# Credit derivatives: instruments of hedging and factors of instability. The example of "Credit Default Swaps" on French reference entities.

Nathalie Rey[*]
CEPN, University of Paris 13

## Abstract

Through a long-period analysis of the inter-temporal relations between the French markets for credit default swaps (CDS), shares and bonds between 2001 and 2008, this article shows how a financial innovation like CDS could heighten financial instability. After describing the operating principles of credit derivatives in general and CDS in particular, we construct two difference VAR models on the series: the share return rates, the variation in bond spreads and the variation in CDS spreads for thirteen French companies, with the aim of bringing to light the relations between these three markets. According to these models, there is indeed an interdependence between the French share, CDS and bond markets, with a strong influence of the share market on the other two. This interdependence increases during periods of tension on the markets (2001-2002, and since the summer of 2007).

*Key words:* credit derivatives, credit risk, credit default swap, inter-temporal relations between markets, VAR models
*JEL classification:* G10; C32

## 1. Introduction

Over the last ten years, with the increase in the number of business bankruptcies and the introduction of a regulatory framework, lending institutions have started actively to manage their credit risk. This has resulted in the rapid development of markets in instruments of risk transfer, including credit derivatives. These latter can be defined as financial contracts reflecting the value of the risk attached to a loan contract. The principle is quite simple. The lending institutions seek to protect themselves against the credit risk, i.e. against the risk of loss incurred in the event of default on the credit. This can be done either by transferring the asset bearing the risk, through what are known as "funded" credit derivatives, or by simply transferring the risk attached to the credit by means of "unfunded" derivatives. In the case of funded derivatives, the underlying risky asset disappears from the balance sheet of the institution buying the protection, while the institution selling the protection buys a bond or debt which then appears in its balance sheet. Where unfunded derivatives are concerned, the underlying asset is not transferred and only the instrument of risk transfer appears, in or outside the balance sheet depending on the instrument used. In practice, however, the use of these instruments raises certain problems and criticisms. These problems are related to the very nature of the contracts, the means of determining the prices of these new products, the lack of transparency of credit derivative markets and the interdependence of these negotiated markets with other markets, including share and bond markets. This article has a double aim. Through an analysis of credit default swaps, the most commonly-used form of credit derivative[†], it will seek to expose both the advantages and the limitations of these instruments of risk transfer. Based on an empiricial study of the inter-temporal relations between the French CDS, share and bond markets over the period 2001-2008, it will show how these derivatives can be the source of financial instability. In the second section, it will situate credit derivatives in relation to other instruments of credit risk transfer, before focusing more particularly on credit default swaps and the literature analysing the relations between the markets for CDS, shares and bonds. The third section presents the credit derivatives market and the empirical data. The econometric method used, the main hypotheses tested and the principal results are presented in the fourth section.

---
[*] E-mail : nathalie.rey@univ-paris13.fr

## 2. Credit derivatives: "simple" instruments of coverage?
### 2.1. Credit derivatives: one of several instruments of risk transfer

Before drawing up a typology of the different instruments of credit risk transfer available to lending institutions and the place of credit derivatives within this typology, it is worth recalling the different forms of underlying credits involved. These can be grouped into two categories: consumer credits (credits outstanding on credit cards, residential real estate loans, loans and financing for the purchase of cars), and securities on transferable and non-transferable loans and debts (customer debts, financing on plant and machinery, commercial mortgages, private sector and sovereign debt). The underlying credit constitutes the primary criterion for the classification of instruments of credit risk transfer. Credit derivatives only concern transferable and non-transferable private sector and sovereign debts, in the form of loans and bonds. To this criterion, we can add that of the modality of transfer of the underlying asset (the funding), depending on whether it is transferred from the buyer of the protection to the seller, or whether only the risk is transferred. Unlike other instruments of risk transfer, the use of credit derivatives does not result in the loaan being removed from the balance sheet of the buyer of the protection, and consequently requires no financing on the part of the seller of the protection.

---

[†] According to the 2006 Fitch survey, the total nominal value of credit derivatives at the end of 2005 stood at nearly 12 trillion dollars, nearly half of which was constituted by CDS.

If a lending institution has sovereign and/or private sector debts among its assets, and if it wishes to keep these assets in its balance sheet, maintaining their legal characteristics and the commercial relation with its customers, while at the same time protecting itself against the risk of default by these customers, then it has the choice between four different instruments, including credit derivatives. These latter correspond to four main types of derivatives, described as "single name" when they cover one sole reference entity (the issuer of the debt for which the lender is seeking coverage), or "portfolio" when there is a basket or portfolio of reference entities (Table 1).

**Table 1 : Credit derivatives**

|  | « Funded » | « Unfunded » |
|---|---|---|
| « single name » | « Credit linked notes » : CLN | « Credit default swap »: CDS<br>« Credit swap option »: CSO<br>« Total swap return »: TSR |
| « portfolio » | CLN | « First-to-default swap »: FTD |

Whatever the type of credit derivative, it is characterised by at least six parameters:
- the initial contract: the underlying loan or asset that the lending institution is seeking to cover, characterised essentially by an amount, a maturity and a rate;
- the coverage contract: the derivative enabling the institution to protect itself against the credit risk, characterised by an amount covered, a maturity and a premium called the spread;
- the buyer of the protection or seller of the risk: two terms defining the institution which can present the initial loaan in its assets and which can use the credit derivative as an instrument of credit risk coverage;
- the reference entity or borrower deemed by the lending institution to represent a credit risk;
- the seller of the protection or buyer of the risk: the institution that undertakes to pay a certain sum in the case of a "credit event" on the reference entity. In the case of the seller of protection, the credit derivative is used as an instrument of investment, as it is buying a risk from the counterparty without having to fund and establish a commercial relation with this latter;
- the credit event corresponding to the occurrence of the credit risk; this may take different forms such as the bankruptcy of the reference entity, default on payment, or restructuring of its debt.

Most of these parameters must be defined within the frame of a contract drawn up between the buyer and seller of the protection. Credit derivative contracts have the particularity that they can be negotiated separately from the underlying asset. Because of this, they can be used as trading instruments, in other words with the sole aim of producing short-term value-added through the hoped-for evolution of the credit risk.

## *2.2. CDS: "first" among credit derivatives*[‡]

A CDS is a contract by which the buyer of protection (A) pays a regular premium (usually every quarter) to the seller (C), who undertakes to compensate (A) for the loss incurred on the underlying asset if a credit event takes place in the reference entity (B) (Figure 1).

**Figure 1:**

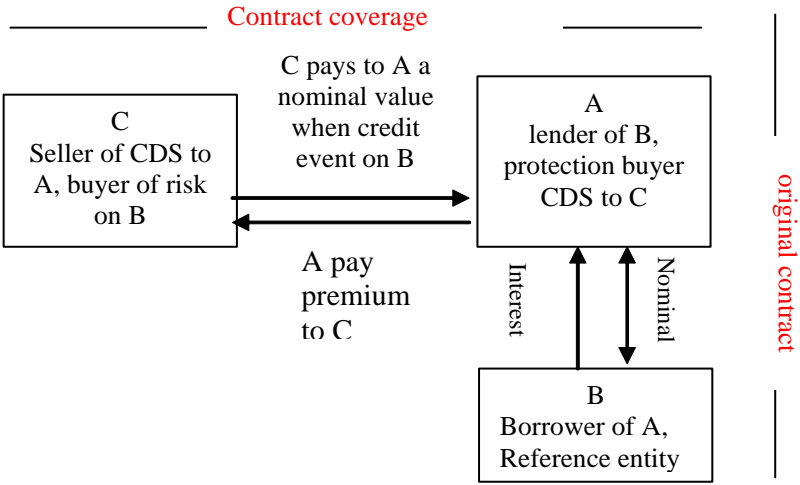

The CDS has no effect on the conditions of the initial contract; the lending institution (A) receives the interest and repayments of the capital loaned from the reference entity (B) as long as no credit event occurs. The lender (A) is not obliged to inform the borrower (B) of the establishment, with the counterparty (C), of an operation of coverage of the credit risk attached to their commercial relation. The buyer of the protection (A) pays the seller protection (C) a regular premium corresponding to the value of the CDS contract, expressed in "basis points". This annual premium, or its equivalent, is inversely proportional to the credit rating of the borrower (B). By paying this premium, the buyer of the protection acquires the right, if a credit event occurs in the reference entity (B), to receive a payment from the seller of the protection (C), corresponding to the market value at the date of establishment of the CDS contract on the debts issued by a borrower with the same credit rating.

---

[‡] For a presentation of the different instruments of risk transfer and other credit derivatives, see notably BATTEN J., HOGAN W. (2002); BIS (2003); KIFF J., MICHAUD F. L. and MITCHELL J. (2003).

Rather than swaps, CDS correspond to put options on the credit risk of an underlying asset. In return for the payment of a premium, the buyer of the CDS has the right to "exercise" this protection in the case of a credit event on the underlying asset, and to receive payment from the protection seller of a sum defined in the CDS. In the case of standard options, however, the buyer's profit (the seller's loss) can be illimited. With CDS, when a credit event occurs, the one's profit and the other's loss are fixed, corresponding to the amount of the indemnity minus the premium. If, on maturity of a CDS contract (which need not correspond to the maturity of the underlying asset covered), the initial borrower (B) has not defaulted, then the protection seller (C) makes a profit equal to the total premium received, and the protection buyer (A) makes a loss equal to the total premium paid minus the reduction in regulatory capital requirements. For the intial lender (A), CDS has a double interest: not only can the lender transfer the credit risk, it can also reduce its capital requirements, by declaring as counterparty of the underlying asset not the initial borrower (B), but the protection seller (C), less risky and with a better credit rating. CDS also has a double interest for the protection seller, in that it makes a profit and diversifies its portfolio.

Let us take a concrete example: that of CDS on the entity France Telecom. On 08/02/2007 the five year CDS spread was 23.5 basis points, meaning that the protection buyer who, for example, wished to cover 10 million euros, would have to pay 5.875 basis points, i.e. a premium of 5,875 euros every quarter over five years: a total of 117,500 euros. If a credit event was to occur in the reference entity France Télécom before 08/02/2012, the protection seller would pay the buyer the total covered (10 million euros) minus the amount recovered from liquidation of the entity, that is to say, with a recovery rate of 40%, six million euros. On the other hand, if the entity France Telecom has not defaulted before 08/02/2012, then the protection seller will have paid nothing during the whole lifespan of the CDS. The CDS spread is an indicator of the market valuation of the credit risk on an entity, and the evolution of this spread gives us information about the market's perception of this risk. Thus, for France Telecom, the five-year CDS spread was 730 basis points on 26 June 2002. The buyer of this contract therefore undertook to pay 182,500 euros every quarter during five years (i.e. a total of 3,650,000 euros!). Between 26 June 2002 and 08/02/2007, the market therefore strongly downgraded the credit risk for France Telecom.

CDS contracts can be negotiated independently of the underlying asset. In this case, the protection buyer (A) has no lending contract with the entity (B), but it establishes a CDS contract on (B) with the protection seller (C) with the simple aim of making a profit. This is the profit it can make by re-selling the CDS contract at a higher price, in the event of an increase in the credit risk on the entity (B). CDS can be used as trading instruments, with the sole objective of producing short-term added-value on the predicted evolution of the credit risk. The buyer of the CDS expects a rise in the credit risk, while the buyer expects a fall.

The buyer and seller of protection must be able to evaluate the credit risk, to attribute a price to it. To determine this price or spread, they use credit risk assessment models. Two types of model are generally opposed: structural models (MERTON, 1974) and reduced-form models (JARROW and TURNBULL, 1995 and 2000; HULL and WHITE, 2000 and 2001). They two categories of models differ on a number of points. In structural models, the financial structure of the entity (the weight of the debt) is taken into consideration, the value of the entity's assets is modelled, and the default can be predicted. In reduced-form models, the fundamentals of the entity are ignored, only the default time is modelled, without any reference to the entity's assets, and the default is totally unpredictable. Structural models enable the spread to be assessed on the basis of fundamental variables, while reduced-form models consider the spread as a given. However, whatever the type of model, three essential parameters must be determined in order to measure the credit risk: the probability of default, the loss in the event of default and the correlation between the two. Now, the determination of these three parameters depends on the information available. Accessible, good-quality information allows the default to be predicted, while a lack of information may make it impossible to predict the default, justifying the use of reduced-form modelling.

The reduced-form approach enables us to relate CDS spreads to bond spreads, that is to say, the difference between the bond yield and the yield of a risk-free asset (DUFFIE and SINGLETON, 1999). If we consider a market without arbitrage and a risk-neutral default probability, then there is a relation of equivalence between the two spreads. This means that the risk assessment is the same on the CDS market and the bond market. In a market without arbitrage, we can construct a replication portfolio, which consists in selling short the fixed-rate bond with the same maturity as the CDS and investing in a risk-free asset. The value of this portfolio is equal to the CDS spread on the entity that issued the bonds. If, for a given entity and a given maturity, the CDS spread and the bond spread are not equal, then arbitrage is possible on the markets. When the CDS spread is higher than the bond spread, investors make a profit by selling the CDS on the derivatives market and buying a risk-free asset and the entity's bond on the bond market.

Using an econometric study of the relation of equivalence between CDS and bond spreads, ZHU (2004) showed that over the long term, this relation is verified. The two spreads evolve together over the long term, but there are shoret-term differences. They react differently to changes in the credit quality of the entity. The derivatives market adapts more quickly than the bond market, where the movements lag behind those of the CDS. Several factors might explain a gap between the two spreads. Bonds are influenced by liquidity, by their life-span, which decreases every day, while the CDS market quotes a set of credit events for a entity with a constant due date. Bonds can include particular clauses such as convertibility into securities, the use of collaterals, which directly affects their value. The eventual impossibility of selling short on the bond market justifies a positive gap between the two spreads.

The structural approach enables us to bring to light a relation between the CDS market and the share market. This approach, developed by MERTON (1974), proposes an interpretation of the relations between shareholders and lenders drawn from the theory of options. It considers that the return profile of the shareholder is similar to that of a call option, and the return profile of the creditor is similar to that of a put option. The more the value of the company's assets increases, the greater the profit the shareholder makes from his shares, after repayment of lenders. The value of the company's assets is determined on the basis of an option model and the information contained in the asset price is assumed to contain implicit information on default. For the lender, the value of the assets must be higher than or equal to the debt, if he is to be reimbursed in the event of bankruptcy. The company defaults when the value of its assets falls below a certain threshold, generally situated somewhere near the value of the debt. Assessment of the probability of default therefore depends on future probability distribution of the share price, relative to the level of debt. A fall in the share market leads to a rise in credit spreads, implying a rise in the probability of default and an increase in the demand for protection (through CDS). To honour their commitments, protection sellers take position on the share market by selling short or to close. The initial shock spreads and the liquidity of the share market falls. This shock may spread to other markets if the protection sellers are forced to sell assets other than shares.

From the moment that structural models can be used to determine, for a given entity, the proportions in which the CDS spread should evolve for a given variation in the share price, it is possible to use the share market to cover credit risk. An investor in an entity's bonds can cover his portfolio either by buying protection on the CDS market or by selling short shares in the same company in the proportions indicated by the structural model. If the bonds market deteriorates, the loss on the bonds will be compensated for by the gain from the protection or from the sale of shares. If all bond investors adopt this same covering strategy, the fall in the bond market will accelerate and spread to the other markets.

On a theoretical level, these two approaches bring to light the relations between the three markets: shares, bonds and CDS. In what follows, this article will seek to measure the extent of these relations, through an econometric study.

## 3. Presentation of the credit derivatives market and the data used
### 3.1. A credit derivatives market dominated by CDS

The credit derivatives market is experiencing strong growth[§]. According to surveys by the credit ratings agency Fitch, the total nominal value of credit derivatives grew from 3 trillion dollars in 2003 to 12 trillion dollars in 2005, with CDS contracts representing nearly half of this total. This market is dominated by short-term contracts; more than 80% of contracts have a maturity date of five years or less. Now, an instrument of coverage loses its efficiency when its maturity is lower than that of the underlying loan.

The banks are the main actors in this market, both as buyers and sellers of protection, although they are, overall, net buyers of protection (for a total of 268 billion dollars in 2005). However, some British and Swiss banks have become net sellers of protection. They use credit derivatives more as financial instruments for the diversification of income sources rather than as an instrument of coverage for their asset portfolios. The main net sellers of protection are insurance and reinsurance companies, pension funds and hedge funds, with 20 to 30% of market activity in 2005. The credit risk is thus transferred from the big international banks to insurance companies and funds, but also to regional banks, notably German ones. These regional banks are turning to this market as a means of diversifying their portfolios by regions and sectors of activity. They use credit derivatives as instruments to generate additional income. Today, credit risks are ultimately transferred to financial institutions that have less ability to manage the risk and to deal with the losses incurred when credit events do occur. The credit derivatives market is very concentrated. In 2005, fifteen institutions were responsible for 86% of total volumes[**]. Now, high concentration tends to increase the counterparty risk.

---

[§] The data on this market are widely dispersed. Public data remain insufficient and private data (British Bankers' Association (BBA) and Fitch are the main suppliers of data on this market) are based on questionnaires and should be treated with caution. The figures quoted in this paragraph (3.1.) are drawn from Fitch surveys.

The reference entities of credit derivatives are few in number. In 2005, 62% of them were companies (with General Motors, DaimlerChrysler, Ford Motors and France Telecom far ahead of the others), 18% were financial institutions (Deutsche Bank, Goldman Sachs, JP Morgan) and 4% were States (Brazil, Italy, and Russia). To an ever increasing extent, the credit derivatives market involves the entities with the worst credit ratings and therefore the highest levels of risk[††]. Between 2003 and 2005, the share of entities with a rating of at least "A" fell from 60% to 33%, while the share of "speculative grade" entities (rated below BBB) grew from 20% to 37%.

### 3.2. The data base on the French CDS, share and bond markets

We have chosen thirteen French companies for our study (including 11 in the CAC40), from twelve different sectors of activity and representing a total market capitalization of nearly 600 billion euros on 23/04/07. For each company, and over the period 2001 - February 2008, we have noted the daily share prices and the daily five-year CDS spreads, and we have calculated the five-year bond spreads (from January 2001 to February 2007). This gave us a total of 60,000 items of data (Table 2). Out of the thirteen companies, two, Alcatel and Rhodia were rated as "speculative grade", and both of them saw their rating slip over the period in question. None of the companies were rated triple A and only three were rated double A.

---

[**] The French banks are active on this market, as there are four of them among the top twenty-five operators. In terms of volume, at the end of 2005, BNP Paribas was in 10th position; the Société Générale and Calyon were in 17th and 18th position respectively.

[††] The principle of credit-rating is that the less risky the entity, the higher the rating. For Standard and Poor's, for example, the best rating is triple A (AAA); the worst is D for issuers who have defaulted. It uses the following rating grid: AAA, AA, A, BBB, BB, B, CCC, CC, D. An entity is considered "investment grade" when it is attributed a rating AAA, AA, A or BBB. Entities with ratings under BBB are considered "speculative grade" and present a very high credit risk.

**Table 2: Firm characteristics**

|  | Period | Sector | Market Capitalization (€) the 23/04/07 | | Number off observations (CDS, bonds, stocks) |
|---|---|---|---|---|---|
| Alcatel | 26/06/2001 – 21/02/2008 | Communications Equipment | 21 480 016 011,30 | 3,61 | 4 941 |
| Sanofi-Aventis | 04/09/2001 - 21/02/2008 | Pharmaceutical Industry | 90 959 774 639,53 | 15,28 | 4 849 |
| BNP Paribas | 03/09/2001 - 21/02/2008 | Banks | 79 761 110 895,00 | 13,40 | 4 825 |
| Bouygues | 19/06/2002 - 21/02/2008 | Construction and Related Machinery | 20 043 133 894,21 | 3,37 | 4 428 |
| Carrefour | 04/09/2001 - 21/02/2008 | Grocery Stores | 40 574 200 332,96 | 6,82 | 4 850 |
| Danone | 04/09/2001 - 21/02/2008 | Packaged Foods | 32 396 792 805,74 | 5,44 | 4 607 |
| France Télécom | 29/08/2001 - 21/02/2008 | Integrated Telecommunication Services | 54 713 063 223,00 | 9,19 | 4 811 |
| Pinault PR | 15/03/2002 - 21/02/2008 | Department Stores | 16 314 170 907,18 | 2,74 | 4 404 |
| Rhodia | 25/03/2002 - 21/02/2008 | Specialty Chemicals | 3 588 474 798,52 | 0,60 | 4 516 |
| Renault | 29/08/2001 - 21/02/2008 | Automobile | 26 539 043 170,52 | 4,46 | 4 835 |
| Société Générale | 03/09/2001 - 21/02/2008 | Banks | 69 836 607 458,70 | 11,73 | 4 664 |
| Sodexho | 22/07/2002 - 21/02/2008 | Restaurants | 8 981 811 806,24 | 1,51 | 4 254 |
| Total | 04/09/2001 - 21/02/2008 | Integrated Oil & Gas | 129 927 028 347,90 | 21,83 | 4 643 |
| **Totaux** | | **12** | **595 115 228 290,80** | **100,00** | **60 627** |

The daily share return, denoted RS, is calculated without dividends, and corresponds to a simple variation in price:

$$RS_t = \log\left(\frac{P_t}{P_{t-1}}\right)$$

with P: the share price observed

The daily CDS spread (denoted by CDS) is calculated as the average between the bid and the ask price for the five-year CDS (the most widely-exchanged maturity). This spread is expressed in 100 base index.

The bond spread (denoted by BOND), expressed in 100 base index is obtained by calculating the difference between the actuarial return rate of the five-year bond with the daily interest rate of the five-year swap curve. For each company, we took the actuarial return rate of a fixed-rate, non-convertible, non-puttable and non-callable five-year bond. When such a bond did not exist, either we calculated the actuarial return rate by interpolation between the rate of bonds of more than five years and those of less than five years maturity, or we took the return rate of bonds of at least three and a half years maturity.

From observation of the three variables, we note (Figures 2, 3 and 4):

- Sharp falls in share prices and negative, very volatile share return rates in 2001 and 2002 for all the companies. Rising prices and positive return rates from 2005-2006 up until summer 2007 (the start of the subprime crisis). The thirteen series of share returns appear to be stationary.

- The CDS spreads reached their highest level in 2002. They experienced a period of decline from 2004 before starting to rise again in summer 2007. The gaps between bids and asks fell strongly over the last few years, expressing greater market maturity. The thirteen CDS spreads appear to be non-stationary.

- The bond spreads were very high in 2001, 2002 and 2003 and have been falling since then. In early 2007, out of the thirteen companies, only Rhodia had a return rate above 5%. The thirteen bond spreads appear to be non-stationary.

We have calculated the mean and the volatility for each variable for each year. Three sub-periods stand out:

- 2001-2002: share return rates were mainly negative and very volatile, the CDS and BOND credit spreads were very high and volatile. The companies with the worst credit ratings had the highest and most volatile spreads.

- 2003-2004: the credit spreads showed a downward trend and became less volatile. Overall, share return rates were low, with lower volatility.

**Figure 2 : Share returns (2001-2008)**

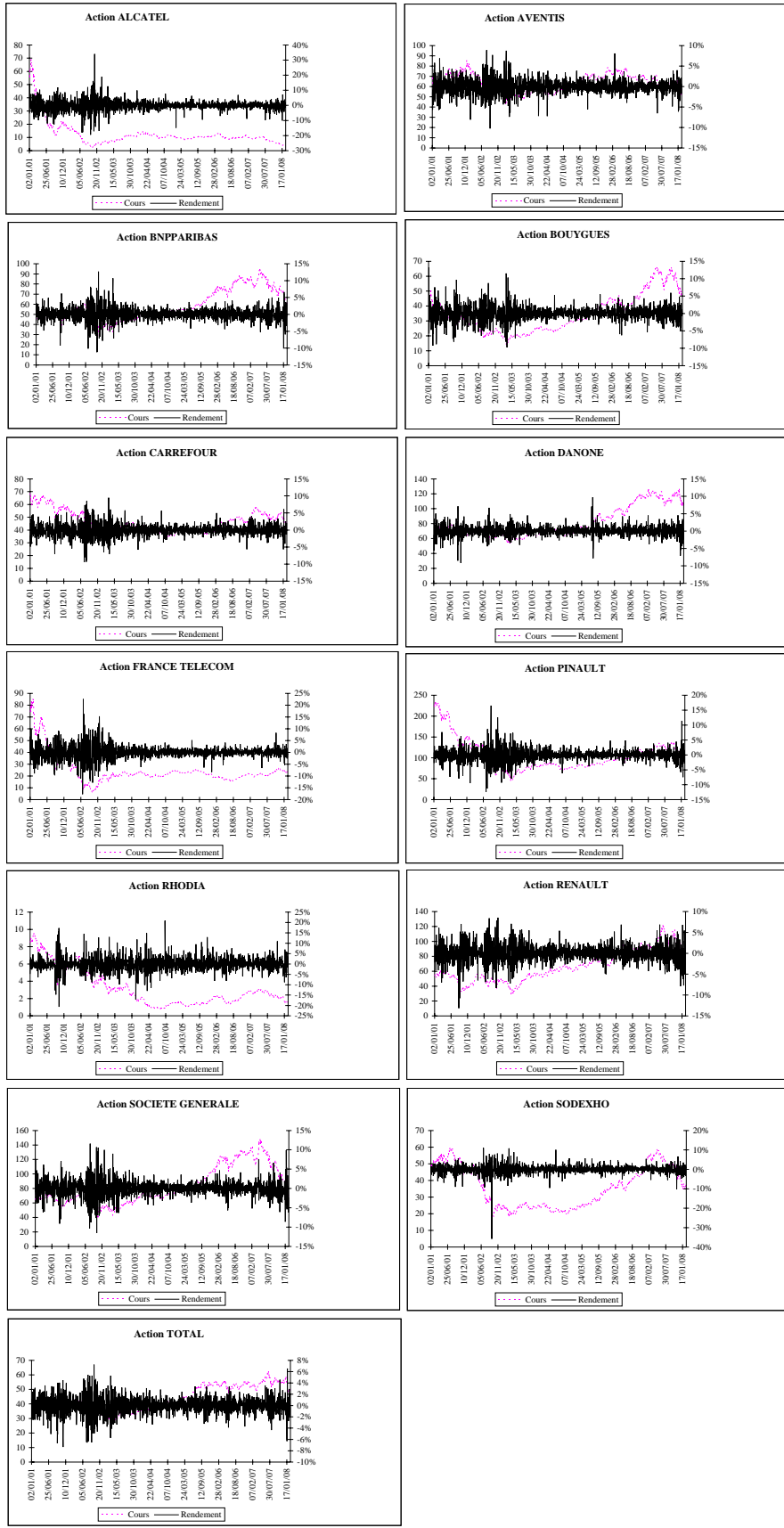

Source : Fininfo

**Figure 3 : CDS spreads (2001-2008)**

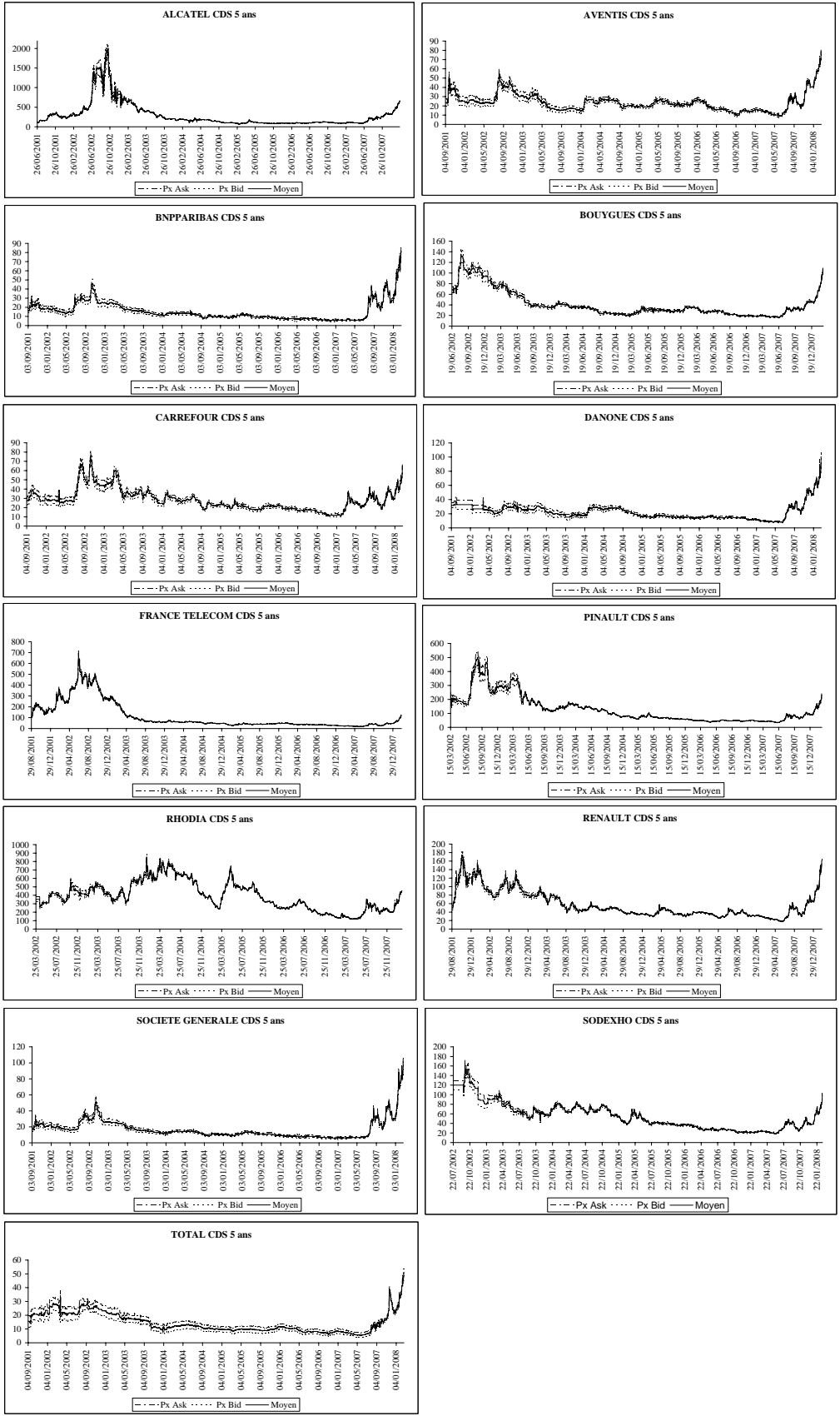

Source : Bloomberg

**Figure 4 : Bonds spreads (2001-2007)**

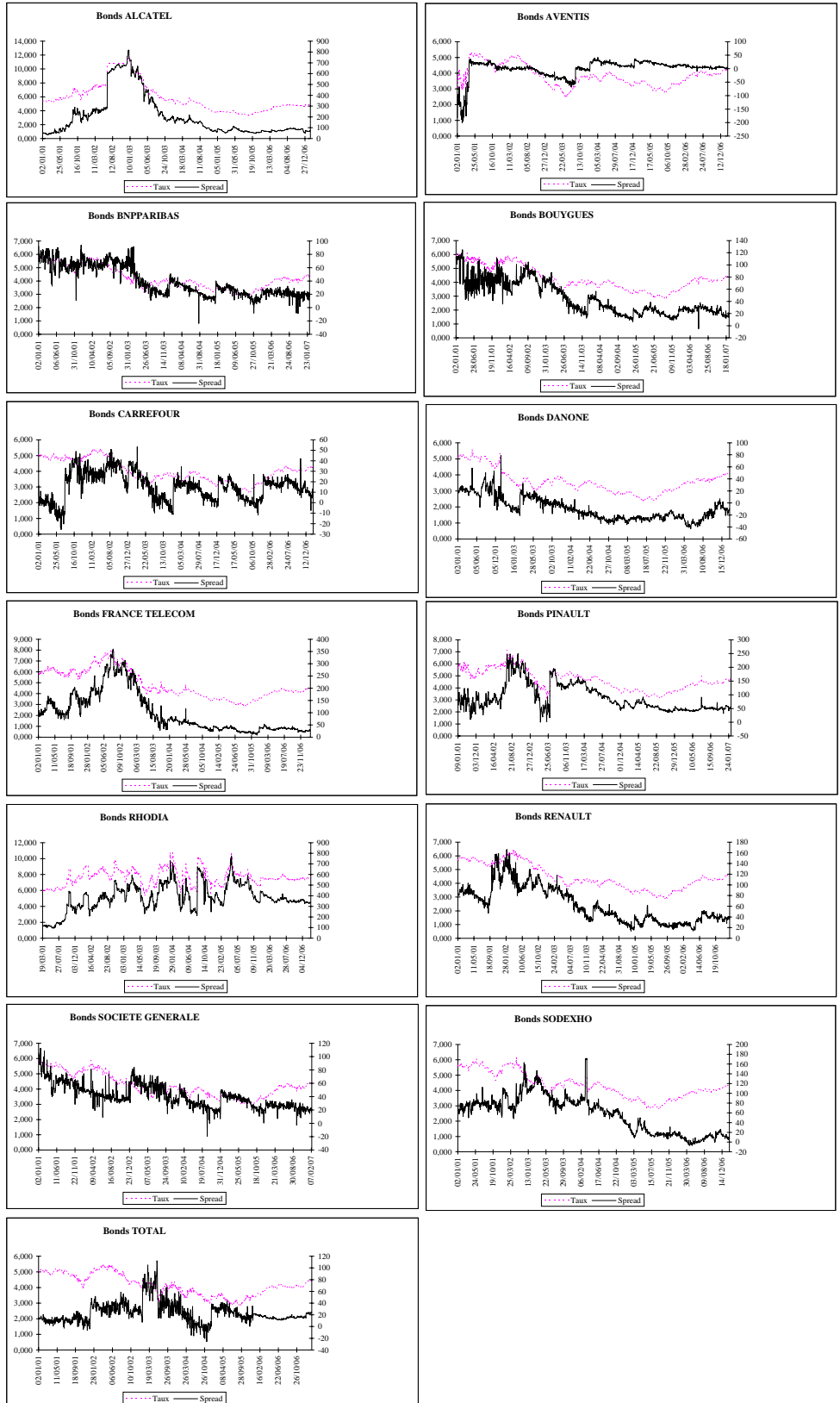

Source : Fininfo

- 2005-2007: overall, share return rates rose, with relatively low volatility, and the credit spreads continued their downward trend. Companies whose ratings had been downgraded saw their spreads increase, while those whose ratings had been upgraded saw their spreads fall.

Since the subprime crisis, there has been negative and volatile average share return rates for the thirteen companies, together with a strong rise in CDS spreads. For six of the thirteen companies, CDS spreads reached their highest levels since 2001. According to the CDS market, the credit risk on these companies from different sectors of activity had all risen sharply. To give one example: from 01/06/2007 to 21/02/2008 the CDS spread on the entities Société Générale and Danone rose from 6.8 to 101.54 and from 8.42 to 100.18 respectively! The losses incurred by the Société Générale may partly explain the strong rise in the spread for this entity, but how can we explain the massive increase in the credit risk assessment for Danone and other companies outside the banking sector? It would appear that this is the consequence of behaviour by investors using the CDS market more to diversify theit portfolios than to cover credit risks.

We have calculated the coefficients of correlation between the three variables for the period 2001-2007 (Table 3). From this matrix of average correlations, we can observe negative correlations between share return rates and bond spreads, and between share return rates and CDS spreads. The correlation between the two credit spreads is positive but weak. This result shows the existence of diversification gains between the three markets and partly explains the behaviour of protection sellers, who see the derivatives market as a source of high returns.

**Table 3: Average correlations 2001-2007**

|                      | Share returns | Bond spread changes | CDS spread changes |
|----------------------|---------------|---------------------|--------------------|
| Share returns        | 1             | -0,0765             | -0,1124            |
| Bond spread changes  | -0,0765       | 1                   | 0,0641             |
| CDS spread changes   | -0,1124       | 0,0641              | 1                  |

## 4. The models and their main results
### *4.1. The VAR models used*
The aim of our study is to analyse the inter-temporal relations between three French markets: shares, bonds and CDS, represented by thirteen French companies. We therefore felt that multivariate modelling would be appropriate, allowing to explain the evolution of one variable on the basis of an extended set of information (the history of the explained variable and the history of the explanatory variables)[‡‡].

We therefore built two difference VAR models, to bring to light the inter-temporal relations between the markets and to verify certan hypotheses:

- Hypothesis 1: positive share return rates are accompanied by negative variations in CDS and bond spreads.
- Hypothesis 2: the share market and the CDS market influence the bond market.
- Hypothesis 3: the relation between the share market and the CDS market is stronger than the relation between the bond market and the share market.
- Hypothesis 4: the relation between the share return rate and the variations in credit spreads depends positively on the company's credit risk. The higher the credit risk, the stronger the relation.
- Hypothesis 5: the relation between the share return rate and the variations in credit spreads depends negatively on the size of the company. The smaller the company (in terms of capitalization), the stronger the relation.

Observation of our series shows that the share-price, CDS-spreads and bond-spreads variables are non-stationary. We therefore conducted stationarity tests, which confirmed the non-stationarity of the series (Table 4). We transformed our variables into share return rates, first-difference or variation of bond spreads and (DBOND) first-difference or variation of CDS spreads (DCDS). The three stationarity tests carried out on the three variables for each company show that the series RS, DBOND and DCDS are indeed stationary. This result is supported by the autocorrelation test for a lag of p = 5 days (Table 5).

**Table 4: Stationarity tests**

|  | Augmented DF Test | Philipps-Perron Test | KPSS test |
|---|---|---|---|
| Share returns | 13 | 13 | 13 |
| Bond spread changes | 13 | 13 | 13* |
| CDS spreads changes | 13 | 13 | 13 |

Number of firms for which the null hypothesis of non-stationary data can be rejected at 5% (*rejected at 1%).

---

[‡‡] Our work is in the line of those carried out by BLANCO, BRENNAN and MARSH (2004), NORDEN and WEBER (2004), and ZHU (2004).

**Table 5: Autocorrelation tests, average period 2001-2007**

|  | Lag 1 | Lag 2 | Lag 3 | Lag 4 | Lag 5 |
|---|---|---|---|---|---|
| Stock returns | 0,0052 | -0,0188 | -0,0292 | 0,0096 | -0,0240 |
| Bond spread changes | -0,1608 | -0,0570 | -0,0175 | -0,0058 | 0,0165 |
| CDS spreads changes | 0,0025 | 0,0320 | 0,0344 | 0,0075 | 0,0335 |

We used the following two VAR models:

- 3 dimensional VAR model (denoted VAR1) for the period 2001 - February 2007;

$$RS_t = a_S + \sum_{p=1}^{5} b_{Sp} RS_{t-p} + \sum_{p=1}^{5} c_{Sp} DBOND_{t-p} + \sum_{p=1}^{5} d_{Sp} DCDS_{t-p} + e_{St}$$

$$DBOND_t = a_B + \sum_{p=1}^{5} b_{Bp} RS_{t-p} + \sum_{p=1}^{5} c_{Bp} DBOND_{t-p} + \sum_{p=1}^{5} d_{Bp} DCDS_{t-p} + e_{Bt}$$

$$DCDS_t = a_C + \sum_{p=1}^{5} b_{Cp} RS_{t-p} + \sum_{p=1}^{5} c_{Cp} DBOND_{t-p} + \sum_{p=1}^{5} d_{Cp} DCDS_{t-p} + e_{Ct}$$

- 2 dimensional VAR model (denoted VAR2) for the period 2001 - February 2008.

$$RS_t = a_S + \sum_{p=1}^{5} b_{Sp} RS_{t-p} + \sum_{p=1}^{5} d_{Sp} DCDS_{t-p} + e_{St}$$

$$DCDS_t = a_C + \sum_{p=1}^{5} b_{Cp} RS_{t-p} + \sum_{p=1}^{5} d_{Cp} DCDS_{t-p} + e_{Ct}$$

*4.2. The results*

For each company, we estimated the two VAR models for the whole period and for two sub-periods (from 01/01/01 to 31/12/03 and from 01/01/04 to 08/02/07 for the VAR1 model; from 01/01/01 to 31/12/03 and from 01/01/04 to 21/02/08 for the VAR2 model).

Over the whole period, and on average for the thirteen companies, the VAR1 model explains and predicts the credit spreads better than the share return rates. The average $R^2$ are higher for the credit-spread variables and the significant coefficients are numerous for these two variables. The coefficients obtained have the expected signs. Overall, share return rates are not very sensitive to variations in bond and CDS spreads; they react negatively to their past changes, and increases in DBOND and DCDS spreads lead to a very weak fall in RS return rates. The DBOND spread variable reacts essentially to its past evolutions, negatively; it falls with a lag when the share return rate rises and it increases weakly with the CDS spread. The DCDS spread variable reacts very negatively to an increase in share return rates, positively to an increase in bond spreads and positively to its past evolutions (Table 6).

**Table 6: VAR1 estimation results over the period 2001-2007**

|  | RS (1) | (2) | DBOND (3) | (4) | DCDS (5) | (6) |
|---|---|---|---|---|---|---|
| RS(-1) | 0,0047 (0,1718) | 4 | -17,0458 (-1,3743) | 4 | -34,1031 (-4,6753) | 10 |
| RS(-2) | -0,0207 (-0,7219) | 2 | -9,9410 (-0,7181) | 2 | -29,5817 (-3,5251) | 9 |
| RS(-3) | -0,0264 (-0,9164) | 3 | -2,6326 (-0,0011) | 2 | -9,7021 (-2,4597) | 8 |
| RS(-4) | 0,0159 (0,5242) | 4 | -3,0859 (0,0320) | 1 | -24,9162 (-2,5052) | 6 |
| RS(-5) | -0,0201 (-0,6998) | 1 | 12,8331 (1,2435) | 6 | 2,5084 (0,2555) | 2 |
| DBOND(-1) | 0,0000 (-0,5144) | 2 | -0,4262 (-15,0158) | 13 | 0,0196 (0,4837) | 3 |
| DBOND(-2) | 0,0000 (-0,1031) | 2 | -0,2745 (-8,6074) | 13 | 0,0073 (0,2737) | 4 |
| DBOND(-3) | 0,0000 (0,3662) | 2 | -0,1837 (-5,6549) | 12 | 0,0181 (0,5625) | 2 |
| DBOND(-4) | 0,0000 (0,3139) | 1 | -0,1129 (-3,5586) | 9 | 0,0096 (0,4231) | 2 |
| DBOND(-5) | 0,0000 (-0,5842) | 2 | -0,0415 (-1,4929) | 3 | 0,0016 (0,1497) | 2 |
| DCDS(-1) | 0,0006 (0,9458) | 3 | 0,0980 (0,8936) | 5 | -0,0538 (-1,8801) | 6 |
| DCDS(-2) | 0,0003 (0,0765) | 3 | 0,0332 (0,7522) | 4 | -0,0202 (-0,6175) | 8 |
| DCDS(-3) | 0,0003 (0,2520) | 2 | 0,0244 (1,3224) | 6 | 0,0036 (0,2024) | 6 |
| DCDS(-4) | 0,0004 (1,2430) | 2 | 0,0134 (0,5118) | 3 | -0,0174 (-0,5510) | 3 |
| DCDS(-5) | 0,0004 (0,9333) | 3 | 0,1451 (0,8609) | 4 | 0,0460 (1,6601) | 7 |
| Const. (a) | 0,0002 (0,4691) |  | -0,0505 (-0,3136) |  | -0,0827 (-0,4476) |  |
| R-squared | 0,0265 |  | 0,1915 |  | 0,1044 |  |
| Adj, R-squared | 0,0146 |  | 0,1816 |  | 0,0935 |  |
| F-statistic | 2,2278 |  | 21,1609 |  | 9,7340 |  |
| Log likelihood | 3080,9115 |  | -4282,0556 |  | -2826,7628 |  |
| Akaike AIC | -4,9032 |  | 6,8487 |  | 4,4798 |  |
| Schwarz SC | -4,8374 |  | 6,9145 |  | 4,5456 |  |
| Mean dependent | 0,0001 |  | -0,0222 |  | -0,0478 |  |
| S,D, dependent | 0,0218 |  | 9,3616 |  | 6,3449 |  |

Columns (1), (3), (5) represent mean coefficients. In brackets, we report the mean t-student. Columns (2), (4), (6) report the number of firms for which the coefficient of the explanatory variable is significantly different from zero at the 5% level.

Over the sub-period January 2001 - December 2003, the results of the VAR1 model are better. The average $R^2$ of the model are higher than those for the whole period. The model explains better a period of declining share market and high, volatile credit spreads. The relations between the three markets are stronger during a period of market tension. There may, therefore, be a risk of contagion between the three markets. During a period when the share market is rising, on the other hand, the relation between the three markets tends to weaken (Table 7).

**Table 7: VAR1 estimation results over the period 2001-2003**

|         | RS (1)              | (2) | DBOND (3)            | (4) | DCDS (5)              | (6) |
|---------|---------------------|-----|----------------------|-----|-----------------------|-----|
| RS(-1)  | 0,0143 (0,3289)     | 2   | -20,5233 (-1,1429)   | 4   | -40,7411 (-3,3373)    | 9   |
| RS(-2)  | -0,0218 (-0,4596)   | 0   | -16,6965 (-0,8500)   | 1   | -37,3088 (-2,7262)    | 8   |
| RS(-3)  | -0,0269 (-0,5765)   | 1   | -2,0749 (-0,1405)    | 1   | -13,5768 (-1,9093)    | 6   |
| RS(-4)  | 0,0244 (0,4909)     | 3   | -3,3614 (-0,1255)    | 1   | -28,5470 (-1,7941)    | 5   |
| RS(-5)  | -0,0278 (-0,6021)   | 1   | 17,6284 (1,0188)     | 1   | -1,0238 (0,0116)      | 1   |
| DBOND(-1) | 0,0000 (-0,4130)  | 0   | -0,4453 (-9,6624)    | 13  | 0,0138 (0,0331)       | 1   |
| DBOND(-2) | 0,0000 (-0,2228)  | 0   | -0,2914 (-5,5713)    | 11  | 0,0103 (0,2914)       | 2   |
| DBOND(-3) | 0,0000 (0,1874)   | 2   | -0,2015 (-3,7864)    | 10  | 0,0232 (0,4995)       | 1   |
| DBOND(-4) | 0,0000 (0,2167)   | 1   | -0,1314 (-2,5183)    | 7   | 0,0150 (0,3589)       | 2   |
| DBOND(-5) | -0,0001 (-0,3997) | 1   | -0,0420 (-0,9499)    | 2   | 0,0037 (0,0930)       | 1   |
| DCDS(-1)  | 0,0007 (0,7611)   | 3   | 0,0906 (0,4365)      | 2   | -0,0884 (-1,8004)     | 6   |
| DCDS(-2)  | 0,0004 (0,3474)   | 1   | 0,0042 (0,5249)      | 2   | -0,0263 (-0,5090)     | 4   |
| DCDS(-3)  | 0,0004 (0,1924)   | 0   | 0,0211 (1,0207)      | 3   | -0,0031 (0,0599)      | 7   |
| DCDS(-4)  | 0,0004 (0,9100)   | 2   | 0,0290 (0,5476)      | 2   | -0,0137 (-0,2518)     | 3   |
| DCDS(-5)  | 0,0005 (0,7192)   | 1   | 0,1714 (0,5777)      | 3   | 0,0480 (1,0930)       | 6   |
| Const. (a) | -0,0003 (-0,1190) |    | -0,0756 (-0,3287)    |     | -0,1451 (-0,4031)     |     |
| R-squared      | 0,0428     |   | 0,2190     |   | 0,1399     |   |
| Adj, R-squared | 0,0109     |   | 0,1931     |   | 0,1114     |   |
| F-statistic    | 1,3422     |   | 9,5659     |   | 5,0390     |   |
| Log likelihood | 1065,1197  |   | -1788,3300 |   | -1281,4857 |   |
| Akaike AIC     | -4,3044    |   | 7,3390     |   | 5,2688     |   |
| Schwarz SC     | -4,1631    |   | 7,4803     |   | 5,4101     |   |
| Mean dependent | -0,0004    |   | 0,0188     |   | 0,0087     |   |
| S,D, dependent | 0,0293     |   | 11,5342    |   | 9,1311     |   |

Columns (1), (3), (5) represent mean coefficients. In brackets, we report the mean t-student. Columns (2), (4), (6) report the number of firms for which the coefficient of the explanatory variable is significantly different from zero at the 5% level.

For the sub-period 2004 - February 2007, the average $R^2$ obtained are lower and the coefficients are less significant. For this sub-period, however, we observe that the credit spread variables explain the share return rate variable better than they do over the whole period (with the expected negative sign) and the bond spreads appear to react less to the share return rate and more to CDS spreads (with the expected positive sign). Finally, the CDS spreads react less to increases in share return rates (with the expected negative sign) than to their own past evolutions (Table 8).

**Table 8: VAR1 estimation results over the period 2004-2007**

|  | RS (1) | (2) | DBOND (3) | (4) | DCDS (5) | (6) |
|---|---|---|---|---|---|---|
| RS(-1) | -0,0225 (-0,6074) | 1 | -12,2426 (-0,6805) | 2 | -8,9482 (-1,5527) | 5 |
| RS(-2) | -0,0192 (-0,5137) | 1 | 1,0875 (0,2959) | 0 | -4,6657 (-0,4390) | 1 |
| RS(-3) | -0,0239 (-0,6657) | 2 | -1,6128 (0,2015) | 1 | -0,6615 (0,1389) | 1 |
| RS(-4) | -0,0033 (-0,0933) | 1 | 3,2225 (0,5618) | 0 | -3,8490 (-0,4020) | 1 |
| RS(-5) | -0,0090 (-0,2347) | 0 | 5,2822 (0,4607) | 0 | 1,9936 (-0,1578) | 2 |
| DBOND(-1) | -0,0001 (-0,5506) | 3 | -0,4814 (-13,0205) | 13 | 0,0157 (0,8276) | 2 |
| DBOND(-2) | 0,0000 (-0,1882) | 3 | -0,3283 (-7,8160) | 13 | 0,0080 (0,3509) | 2 |
| DBOND(-3) | 0,0000 (-0,0989) | 1 | -0,2103 (-4,8521) | 12 | 0,0075 (0,2236) | 2 |
| DBOND(-4) | -0,0001 (-0,5279) | 1 | -0,1192 (-2,8616) | 10 | 0,0044 (0,3554) | 0 |
| DBOND(-5) | -0,0001 (-0,7223) | 2 | -0,0520 (-1,4108) | 2 | 0,0084 (0,6590) | 1 |
| DCDS(-1) | 0,0004 (0,4340) | 0 | 0,2337 (2,0010) | 5 | 0,0353 (0,9499) | 13 |
| DCDS(-2) | -0,0006 (-0,8405) | 2 | 0,2593 (0,6450) | 2 | -0,0334 (-0,8514) | 5 |
| DCDS(-3) | -0,0001 (0,1561) | 0 | 0,1991 (1,1296) | 2 | 0,0229 (0,6557) | 4 |
| DCDS(-4) | 0,0005 (0,6229) | 0 | 0,1121 (0,7927) | 1 | -0,0009 (-0,0075) | 2 |
| DCDS(-5) | 0,0002 (0,4894) | 2 | 0,0211 (0,5253) | 2 | 0,0339 (0,9372) | 3 |
| Const. (a) | 0,0005 (0,9891) |  | -0,0224 (-0,1653) |  | -0,0691 (-0,8092) |  |
| R-squared | 0,0254 |  | 0,2140 |  | 0,0953 |  |
| Adj, R-squared | 0,0058 |  | 0,1982 |  | 0,0771 |  |
| F-statistic | 1,3157 |  | 14,0721 |  | 5,3888 |  |
| Log likelihood | 2192,8224 |  | -2309,4307 |  | -936,5150 |  |
| Akaike AIC | -5,7117 |  | 6,0967 |  | 2,4292 |  |
| Schwarz SC | -5,6143 |  | 6,1942 |  | 2,5267 |  |
| Mean dependent | 0,0004 |  | -0,0455 |  | -0,0843 |  |
| S,D, dependent | 0,0145 |  | 6,9592 |  | 2,1017 |  |

Columns (1), (3), (5) represent mean coefficients. In brackets, we report the mean t-student. Columns (2), (4), (6) report the number of firms for which the coefficient of the explanatory variable is significantly different from zero at the 5% level.

The relations that appear to dominate are those of the share market towards the CDS market, the share market towards the bond market and the CDS market towards the bond market.

To test the consequences of the current subprime crisis on the relations of interdependence between the share market and the CDS market, we estimated the VAR2 model. Two main conclusions can be drawn from our results. Firstly, the bond market appears to be relevant for explaining the evolutions of the other two markets, especially the CDS market (for the three periods, the average $R^2$ obtained for the DCDS equation of the VAR2 model are lower than those of the VAR1 model). Secondly, the relations of interdependence between the markets are stronger during periods of financial instability. The VAR2 model explains better the unstable sub-period 2001-2003 (average $R^2$ higher than for the whole period and average $R^2$ of the DCDS equation close to those obtained with the VAR1 model) and show a strengthening of the relation between the share returns and CDS spreads of the companies most affected by the crisis (the $R^2$ of the RS and DCDS equations are higher over the reccent period 2004-2008 that for the whole period for the Société Générale, BNP Paribas, Danone, Renault, and Total). For the period 2004-2008, we obtain higher and more significant average coefficients and the share return rates appear to react more strongly to variations in CDS spreads than to their own past evolutions (Table 9).

We have verified the relations of causality between these three markets over the whole period using the Granger test. For eleven of the thirteen companies, there is a relation of causality from share return rates to CDS spreads, and this causality is bidirectional for eight of the eleven companies. For eight out of thirteen companies, there is a relation of causality from CDS spreads to bond spreads, and this causality is bidirectional for six of the eight companies. For the companies with high credit risk, and therefore with low ratings, there is a relation of causality between their share return, their bond spread and their CDS spread. On the contrary, there is no relation of causality between these markets for the company with the best credit rating, which is also a company with large market capitalization (Table 10).

**Table 9: VAR2 estimation results over the period 2004-2008**

|        | RS (1)              | (2) | DCDS (3)            | (4) |
|--------|---------------------|-----|---------------------|-----|
| RS(-1) | -0,0115 (-0,3657)   | 0   | -12,4042 (-2,4983)  | 9   |
| RS(-2) | 0,0036 (0,1103)     | 0   | -9,0604 (-1,4831)   | 5   |
| RS(-3) | -0,0427 (-1,3535)   | 3   | -1,8805 (-0,1164)   | 1   |
| RS(-4) | 0,0022 (0,0640)     | 2   | -3,1933 (-0,7997)   | 0   |
| RS(-5) | -0,0267 (-0,8304)   | 1   | 4,8128 (0,6724)     | 1   |
| DCDS(-1) | 0,0004 (1,1957)   | 4   | 0,0947 (2,9724)     | 10  |
| DCDS(-2) | -0,0002 (-0,7676) | 3   | 0,0075 (0,2691)     | 6   |
| DCDS(-3) | -0,0004 (-0,9574) | 2   | 0,0921 (2,7894)     | 9   |
| DCDS(-4) | 0,0000 (-0,0505)  | 0   | -0,0388 (-1,1446)   | 4   |
| DCDS(-5) | -0,0001 (0,0626)  | 3   | 0,0043 (0,1493)     | 3   |
| Const. (a) | 0,0001 (0,3929) |     | 0,0626 (1,3870)     |     |
| R-squared | 0,0180           |     | 0,0616              |     |
| Adj, R-squared | 0,0086      |     | 0,0527              |     |
| F-statistic | 1,9330          |     | 7,0034              |     |
| Log likelihood | 2949,4742    |     | -2033,6641          |     |
| Akaike AIC | -5,5286          |     | 3,8470              |     |
| Schwarz SC | -5,4772          |     | 3,8984              |     |
| Mean dependent | 0,0001       |     | 0,0745              |     |
| S,D, dependent | 0,0157       |     | 2,8049              |     |

Columns (1), (3) represent mean coefficients. In brackets, we report the mean t-student. Columns (2), (4) report the number of firms for which the coefficient of the explanatory variable is significantly different from zero at the 5% level.

We have sought to verify whether the current crisis has weakened the relations of causality between the share market and the CDS market. Our results show that this crisis has led to an increase in the number of causalities. According to the VAR1 model and for the sub-period 2004-2007, there is a relation of causality from share returns to CDS spreads for four companies; when we introduce the crisis (VAR2 model) this relation is verified for nine companies: the previous ones plus those in the financial sector and the big groups. Before the crisis, the relation of causality from CDS spreads to share returns existed for one sole company; since the crisis, this relation concerns four companies (Table 11).

**Table 10: Granger causality Tests (VAR1 model)**

|  | DBOND cause RS | RS cause DBOND | DCDS cause RS | RS cause DCDS | DCDS cause DBOND | DBOND cause DCDS |
|---|---|---|---|---|---|---|
| Alcatel | yes | yes | yes | yes | yes | yes |
| Sanofi-Aventis | no | no | yes | yes | no | no |
| BNP Paribas | no | no | yes | yes | no | no |
| Bouygues | no | no | no | yes | yes | yes |
| Carrefour | no | no | no | yes | no | yes |
| Danone | no | no | yes | no | yes | yes |
| France Télécom | no | yes | no | yes | yes | no |
| Pinault PR | yes | yes | yes | yes | yes | yes |
| Rhodia | no | yes | yes | yes | yes | no |
| Renault | no | yes | no | yes | yes | yes |
| Société Générale | no | no | yes | yes | yes | no |
| Sodexho | no | no | yes | yes | yes | no |
| Total | no | no | no | no | no | no |
| **TOTAL** | **2** | **5** | **8** | **11** | **9** | **6** |

**Table 11: Granger causality Tests (comparison of the VAR models results)**

|  | Global period* | | Sub period 1** | | Sub period 2*** | |
|---|---|---|---|---|---|---|
|  | RS cause DCDS | DCDS cause RS | RS cause DCDS | DCDS cause RS | RS cause DCDS | DCDS cause RS |
| VAR1 model | 11 | 8 | 10 | 2 | 4 | 1 |
| VAR2 model | 11 | 6 | 11 | 3 | 9 | 4 |

*: 2001-2007 for VAR1 and 2001-2008 for VAR2; **: 2001-2003 for VAR1 and VAR2; ***: 2004-2007 for VAR1 and 2004-2008 for VAR2.

As our results highlight the existence of a relation of causality between share returns and credit spreads, we thought it appropriate to study the effet of a share market shock of the CDS and bond markets (the three markets being represented by our thirteen companies). We expected an instant rise in credit spreads following a fall in share return rates, and conversely a rise in share return rates should cause an instant fall in spreads. Figures 5, 6 and 7 display the impulse response functions: how do credit spreads react to a share return rate shock?[§§]. We consider that the amplitude of the shock is equal to the standard deviation, and we observe the effects of the shock over fifteen periods, i.e. fifteen days[***]. The share return rate shock has an immediate impact on CDS and bond spreads and the consequences of the shock disappear after seven days. The reaction of the CDS market to a share return rate shock is stronger and lasts longer than the reaction of the bond market. According to the impulse response functions obtained using the VAR2 model applied to the sub-period 2004-2008, the CDS spreads react immediately to a share return rate shock, with a fall in the share market causing an instant rise in spreads on the CDS market. This rise is absorbed in almost linear fashion in less than six days.

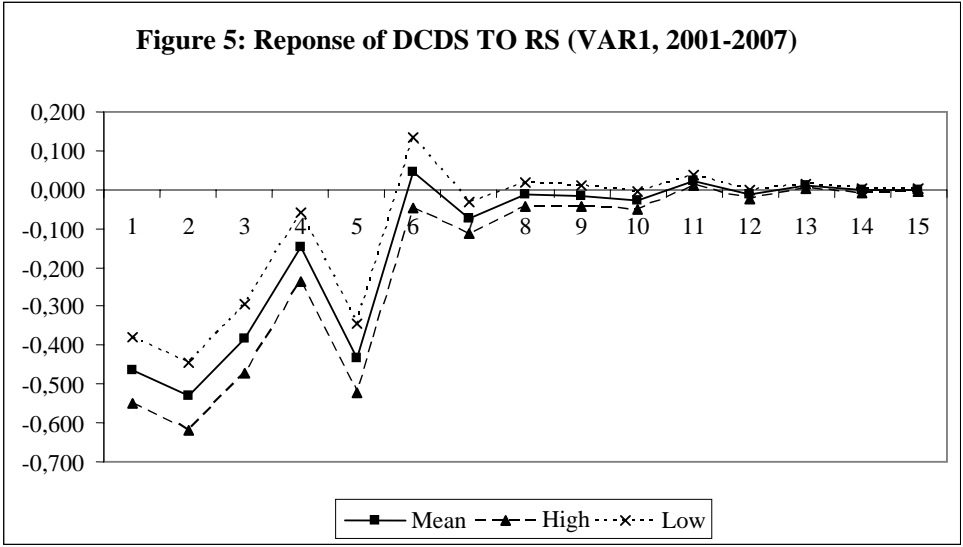

---

[§§] We estimated the impulse response function for each company and then weighted our results by the ratio of the market capitalization of the company to the total capitalization of the thirteen companies to obtain figures 5, 6 and 7.
[***] We used the Cholesky decomposition.

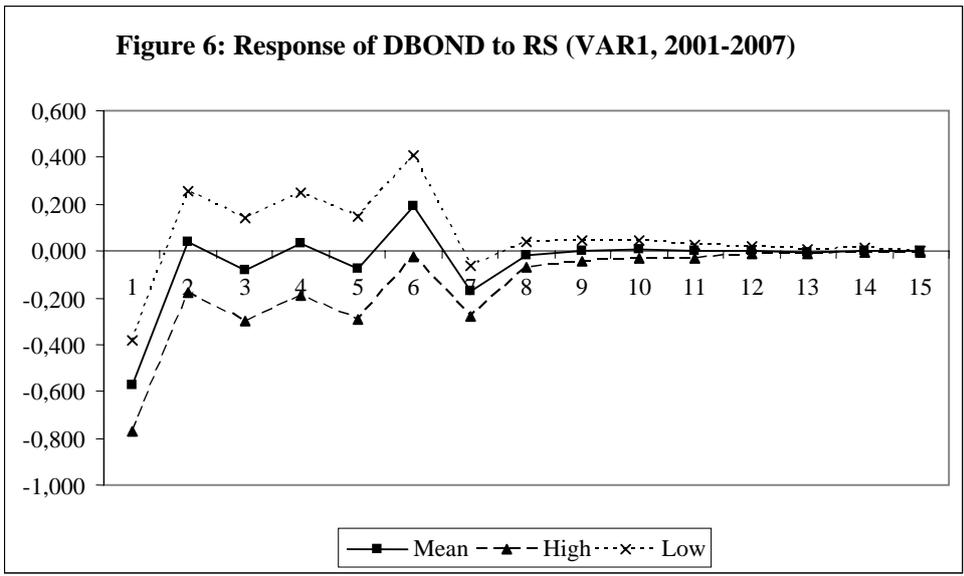

Figure 6: Response of DBOND to RS (VAR1, 2001-2007)

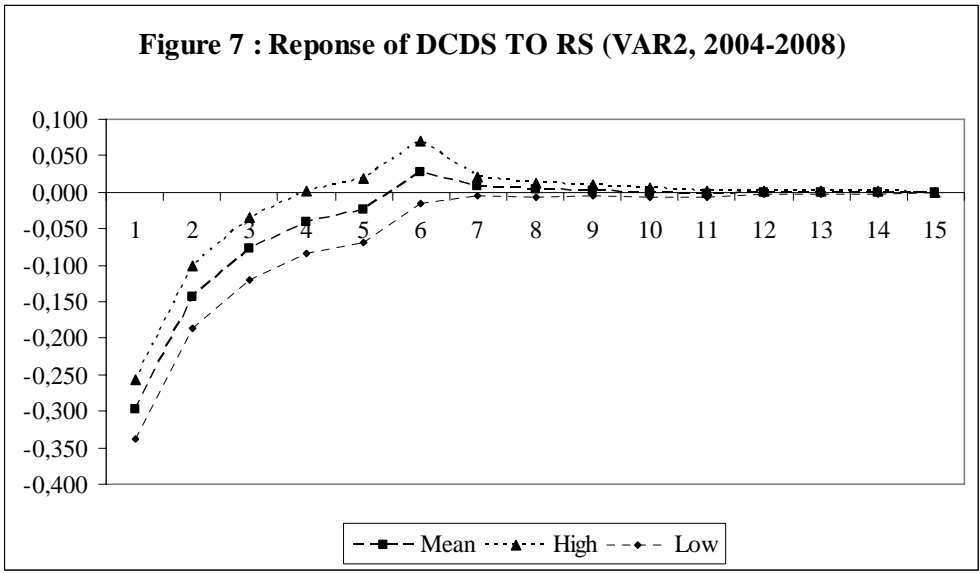

Figure 7 : Reponse of DCDS TO RS (VAR2, 2004-2008)

## 5. Conclusion

The credit derivatives market has developed rapidly over the last ten years. The credit default swaps that dominate this market are used as instruments of credit risk coverage, but they are also used by investors for portfolio diversification. Investors adopt strategies of coverage, behaviours of arbitrage between the markets that are conducive to the propagation of instability between markets. The theoretical models of credit risk assessment bring to light relations between the share, bond and CDS markets. In this article, we have sought to conduct a long-period analysis of the relations between the French markets between 2001 and 2008. Analysis of the survey data and the results of the VAR models used both suggest that the three markets are interconnected, and that this interconnection is all the stronger when the markets experience tensions. The influence of the French share market on the CDS and bond markets is proven, as is the influence of the CDS market on the bond market. During a period of tension on the share market, this interconnection between the markets can favour financial instability.